\documentstyle[preprint,prc,aps,psfig]{revtex}

\renewcommand{\vec}[1]{\mbox{\boldmath $#1$}}

\tightenlines

\begin{document}
\title{Continuum QRPA in the coordinate space representation} 
\author{K. Hagino$^1$ and H. Sagawa$^2$}
\address{$^1$ Yukawa Institute for Theoretical Physics, Kyoto
University, Kyoto 606-8502, Japan \\
$^2$ Center for Mathematical Sciences, University of Aizu, Ikki-machi, \\
Aizu-Wakamatsu, Fukushima 965-8580, Japan}

\maketitle

\begin{abstract}

We formulate a  quasi-particle random phase approximation (QRPA) in 
the coordinate space representation. This model  is a natural extension of
the RPA model  of Shlomo and Bertsch  to open-shell nuclei in order to
take into account pairing correlations together with the coupling to
the continuum.  
We apply it to the $^{120}$Sn nucleus and show that low-lying excitation
 modes  are significantly influenced by the pairing effects
although the effects are marginal in the giant resonance region. 
The dependence of the pairing effect on the parity of low-lying
collective mode is also discussed. 

\end{abstract}

\section{Introduction}

The random phase approximation (RPA) has provided a convenient and 
useful method to describe excited states of many-fermion systems. 
There are a number of ways to formulate the RPA 
\cite{BB94,FW71,R68,RS80,SB75,RG92,BT75,Tsai78,GSV98}.  
In practical point of view, we particularly mention 
here a  configuration space formalism and a  response function formalism. 
The configuration space formalism diagonalizes a non-hermitian matrix denoted 
often by $A$ and $B$ matrices which 
are constructed in the model  space of 1 particle-1 hole 
(1p-1h) states. 
In contrast, the response function formalism is based on 
the linear response theory and solves a Bethe-Salpeter 
equation for the response function, often in the coordinate 
space \cite{SB75,BT75,Tsai78}. 
The response function formalism can be formulated in various
representations of basis states, and the connection between these 
two methods can
be made by expressing response functions in the configuration 
space \cite{BB94}.

The response function formalism of RPA becomes particularly simple when the
interaction is a zero-range contact force. 
In that case, the excitations to particle
continuum states can be treated exactly by solving  the single-particle 
Green function in the coordinate space. This method was
first developed by Shlomo and Bertsch \cite{SB75} and subsequently applied
to self-consistent calculations of nuclear giant resonances 
with Skyrme interaction by Liu and Van Giai \cite{LG76}. 
Recently it has been extensively used by
Hamamoto, Sagawa, and Zhang to discuss giant resonances of neutron-rich
nuclei, where the continuum effects play an essential role due to a much
lower threshold energy compared with $\beta$-stable nuclei 
\cite{HSZ97,HS99}. 
An extension to the three dimensional space has also been carried out 
recently by Nakatsukasa and Yabana for the 
study of atomic clusters \cite{NY00}.  The coupling between
particle-particle continuum was also studied in $^{11}$Li 
by Bertsch and Esbensen\cite{BE91,EB92}. 

Although it 
has been well known that the pairing correlation plays an important role in 
the ground state of open-shell nuclei, 
 it has been neglected in applying the
response function formalism to describe excited states of atomic
nuclei until very recently \cite{KLLT98}. 
So far a simple  filling
approximation has been employed in open-shell nuclei 
in order to simulate the pairing correlations. 
However, it is important to take into account the 
pairing effects consistently for the study of  excited states
of open-shell nuclei, and thus the quasi-particle RPA
(QRPA) should be used instead of the RPA \CITE{KG00}.
 
The aim of this paper is to generalize the formalism of Shlomo and
Bertsch to the QRPA and discuss the effects of
pairing correlations on excited states of open-shell nuclei. 
In this paper, we study only cases where the
BCS approximation works well and thus all states in the pairing 
active space are bound, leaving out as a future work 
a self-consistent treatment of the pairing effects in the continuum 
using the Hartree-Fock-Bogoliubov + RPA theory
\cite{DNW96,FTTZ00,GGLS00}. 
Our method is closely related to that in Ref. \cite{KLLT98}, 
although details are somewhat different. 
The paper is organized as follows. In  Sec. II, we briefly
review the formalism of Shlomo and Bertsch and extend it to the
QRPA. In Sec. III, we apply the formalism to the isoscalar (IS) monopole,
quadrupole, and octupole modes as well as the isovector (IV) dipole mode of
excitation of the $^{120}$Sn nucleus and discuss the effects of
pairing correlations on the excited states. A  summary is given
 in Sec. IV. 

\section{Linear Response Theory for Open-shell Nuclei}

We begin with the configuration space formalism of the RPA theory and then make
a connection to the response function method. 
The RPA equation can be given in a compact form 
\begin{equation}
\left(\matrix{
A & B \cr
-B^* & -A^* \cr}\right)
\left(\matrix{
X \cr Y \cr}\right)
=\omega
\left(\matrix{
X \cr Y \cr}\right)\, ,
\end{equation}
where $X$ and $Y$ are the forward and the backward amplitudes,
respectively. 
For a residual interaction of delta-type contact force, 
$v_{res}(\vec{r_1},\vec{r_2})=v((r_1+r_2)/2)
\,\delta(\vec{r_1}-\vec{r_2})$, the
matrix elements of $A$ and $B$ for the $L$-multipole mode read
\begin{equation}
A_{ph,p'h'}
-(\epsilon_p-\epsilon_h)\delta_{ph,p'h'}
=B_{ph,p'h'}=I(php'h')\langle j_pl_p ||Y_L|| j_hl_h\rangle
\langle j_{p'}l_{p'} ||Y_L|| j_{h'}l_{h'}\rangle\, \frac{1}{2L+1}\, ,
\label{eq:ab}
\end{equation}
where $p$($h$) denotes a particle (hole) state and 
$\langle j l ||Y_L|| j' l' \rangle$ is the reduced matrix element. 
The radial integral $I$ is given by 
\begin{equation}
I(php'h')=\int^{\infty}_0 \frac{dr}{r^2}\,v(r) 
\phi_p(r)\phi_h(r)\phi_{p'}(r)\phi_{h'}(r), 
\label{eq:i}
\end{equation}
where $\phi(r)$ is a single-particle radial wave function.  
We have absorbed the overall factor $(-1)^L$ in front of the $B$ matrix by 
redefining the sign of the backward $Y$ amplitudes. 

The integral (\ref{eq:i}) can be computed by discretizing it as 
\begin{equation}
I(php'h')\approx \sum_k \frac{\Delta r}{r_k^2}\,v(r_k) 
\phi_p(r_k)\phi_h(r_k)\phi_{p'}(r_k)\phi_{h'}(r_k), 
\label{eq:id}
\end{equation}
where $\Delta r$ is the spacing of the radial coordinate. 
Since the interaction is then given as a sum
of separable form, the RPA frequencies
$\omega$ can be obtained by solving the
generalized RPA dispersion relation 
$det(1-\Pi_0(\omega)\chi)=0$ \cite{GSV98}, 
where the matrices $\Pi_0(\omega)$ and $\chi$ 
are given by 
\begin{eqnarray}
\Pi_0(i,j;\omega)&=&-\sum_{ph}D_{ph}(i)D_{ph}(j)
\left(\frac{1}{\epsilon_p-\epsilon_h-\omega-i\eta}
+\frac{1}{\epsilon_p-\epsilon_h+\omega-i\eta}\right)\, , \label{free}\\
\chi(i,j)&=&\chi(i)\,\delta_{i,j}=\frac{\Delta r}{r_i^2}v(r_i),
\end{eqnarray}
respectively, in the coordinate space representation. Here, $\eta$ is
a infinitesimal real number, and $D_{ph}$ is given by 
\begin{equation}
D_{ph}(i)=
\phi_p(r_i)\phi_h(r_i)
\langle j_pl_p ||Y_L|| j_hl_h\rangle\,\frac{1}{\sqrt{2L+1}}\, .
\end{equation}

Note that $\Pi_0(\omega)$ is nothing but the unperturbed response
function in the linear response theory. 
Here we introduce the RPA response function which obeys a
Bethe-Salpeter equation 
\begin{equation}
\Pi_{RPA}=\Pi_0+\Pi_0\chi\Pi_{RPA}. 
\label{BS}
\end{equation}
This equation can be solved in the coordinate space by matrix
inversion as \cite{BT75}
\begin{equation}
\Pi_{RPA}(i,j;\omega)=\sum_k(1-\Pi_0(\omega)\chi)^{-1}_{i,k}
\Pi_0(k,j;\omega).
\end{equation}
The response of the system to an external field 
$V_{ext}(\vec{r})=V_{ext}(r)Y_{LM}(\hat{\vec{r}})$ is then given by \cite{BT75} 
\begin{equation}
S(\omega)\equiv \sum_f|\langle
f|V_{ext}|0\rangle|^2\delta(E_f-E_0-\omega)
=\frac{1}{\pi}Im\,
\int dr_i \int dr_j V_{ext}(r_i)\Pi_{RPA}(i,j;\omega)
V_{ext}(r_j).
\end{equation}

The exact treatment of the continuum effects can be achieved by
eliminating the sum of the particle states in the free response function
(\ref{free}) using the complete set of the wave function \cite{SB75}. 
This leads to 
\begin{eqnarray}
\Pi_0(i,j;\omega)&=&
-\sum_{h}
\phi_h(r_i)\phi_h(r_j)\sum_{j_pl_p}
\langle j_pl_p ||Y_L|| j_hl_h\rangle^2\,\frac{1}{2L+1} \nonumber \\
&& \times \left\langle r_i\left|
\frac{1}{\hat{h}-\epsilon_h-\omega-i\eta}
+\frac{1}{\hat{h}-\epsilon_h+\omega-i\eta}\right|r_j\right\rangle
\, ,
\label{free2}
\end{eqnarray}
where $\hat{h}$ is the single-particle Hamiltonian, 
 and the single-particle Green function is given by 
\begin{equation}
\left\langle r\left|
\frac{1}{\hat{h}-\epsilon_h\pm \omega-i\eta}\right|r'\right\rangle
=-\frac{2m}{\hbar^2}\frac{u(r_<)w(r_>)}{W}\,.
\label{green}
\end{equation}
Here, $u$ and $w$ are the regular and irregular solutions of the
Hamiltonian $\hat{h}$ at energy $\epsilon_h\mp\omega$, 
and $W$ is the Wronskian given by $W=uw'-wu'$. 

In the  application of the formalism to nuclear systems, we use
the proton-neutron formalism which can properly take into account the
couplings between the IS and IV modes of 
excitation \cite{HSZ97}. 
The Bethe-Salpeter equation given by Eq. (\ref{BS}) is then
generalized to  be 
\begin{equation}
\left(\matrix{
\Pi_{RPA}^{(p)} \cr \Pi_{RPA}^{(n)} \cr}\right)
=\left(\matrix{
\Pi_{0}^{(p)} \cr \Pi_{0}^{(n)} \cr}\right)
+
\left(\matrix{
\Pi_{0}^{(p)}\chi_{pp} & \Pi_{0}^{(p)}\chi_{pn} \cr 
\Pi_{0}^{(n)}\chi_{pn} & \Pi_{0}^{(n)}\chi_{nn} \cr }\right)
\left(\matrix{
\Pi_{RPA}^{(p)} \cr \Pi_{RPA}^{(n)} \cr}\right)\,.
\end{equation}
For simplicity of the notation, we do not present below the
formalism in the two-component form, although we indeed 
use it in the actual
calculations given in the next section. 

In the presence of paring correlations, the elementary excitation is
a two quasi-particle excitation, rather than a particle-hole
excitation. A generalization of the RPA to the QRPA is given in Ref. 
\cite{R68} together with relevant $A$ and $B$  
matrices in the QRPA
equation. A free response function in the QRPA in the configuration
space representation can be constructed in
a similar way to  the RPA, 
\begin{equation}
\Pi_0(i,j;\omega)=-\sum_{\alpha\leq\beta}D_{\alpha\beta}(i)
D_{\alpha\beta}(j)
\left(\frac{1}{E_{\alpha}+E_{\beta}-\omega-i\eta}
+\frac{1}{E_{\alpha}+E_{\beta}+\omega-i\eta}\right)\, ,
\label{freeBCS}
\end{equation}
with
\begin{equation}
D_{\alpha\beta}(i)=
\phi_{\alpha}(r_i)\phi_{\beta}(r_i)
\langle j_{\alpha}l_{\alpha} ||Y_L|| j_{\beta}l_{\beta}\rangle\,
\frac{u_{\alpha}v_{\beta}+(-)^Lv_{\alpha}u_{\beta}}{\sqrt{2L+1}}
(1+\delta_{\alpha,\beta})^{-1/2}\, ,
\end{equation}
where $v_{\alpha}^2$ is the BCS occupation probability and
$u_{\alpha}^2=1-v_{\alpha}^2$. $E_{\alpha}$ is the quasi-particle
energy given by 
$E_{\alpha}=\sqrt{(\epsilon_{\alpha}-\lambda)^2+\Delta_{\alpha}^2}$,
where  
$\lambda$ and $\Delta_{\alpha}$ are  the chemical potential and the pairing
gap, respectively. In the BCS approximation,  
$\phi_{\alpha}$ is an eigenfunction of
the single-particle Hamiltonian $\hat{h}$ with an
eigen-energy $\epsilon_{\alpha}$. Since the quasi-particle energy
$E_{\alpha}$ is not an eigenvalue of $\hat{h}$ in
general, it is
not straightforward to introduce the single-particle Green function 
(\ref{green}) in the QRPA free response function. 
However, when the 
pairing gap is zero 
for states $k$ outside the pairing active space,  
the quasiparticle energy becomes an eigenvalue
of the single-particle Hamiltonian, i.e. $E_k=\epsilon_k-\lambda, v_k=0$, and 
$u_k$=1. We therefore consider separately excitations among states within the
pairing active space and those from the inside to the outside of the
active space. To the latter model space, we apply the same procedure
as the RPA response function.  
The free response function in the BCS approximation (\ref{freeBCS}) 
thus becomes 
\begin{eqnarray}
&&\Pi_0(i,j;\omega)=-\sum_{\alpha\leq\beta}D_{\alpha\beta}(i)
D_{\alpha\beta}(j)
\left(\frac{1}{E_{\alpha}+E_{\beta}-\omega-i\eta}
+\frac{1}{E_{\alpha}+E_{\beta}+\omega-i\eta}\right) \nonumber \\
&&\quad-\sum_{\alpha}
\phi_{\alpha}(r_i)\phi_{\alpha}(r_j)v_{\alpha}^2\sum_{j_kl_k}
\langle j_{\alpha}l_{\alpha} ||Y_L|| j_kl_k\rangle^2\,\frac{1}{2L+1} 
\nonumber \\
&& \qquad\times \left\{\left\langle r_i\left|
\frac{1}{E_{\alpha}+\hat{h}-\lambda-\omega-i\eta}
+\frac{1}{E_{\alpha}+\hat{h}-\lambda+\omega-i\eta}
\right|r_j\right\rangle\right. \nonumber \\
&& \qquad \left.-\sum_{\beta}\delta_{j_k,j_{\beta}}\delta_{l_k,l_{\beta}}
\phi_{\beta}(r_i)\phi_{\beta}(r_j)
\left(\frac{1}{E_{\alpha}+\epsilon_{\beta}-\lambda-\omega-i\eta}
+\frac{1}{E_{\alpha}+\epsilon_{\beta}-\lambda+\omega-i\eta}\right)\right\}\, ,
\label{freeBCS2}
\end{eqnarray}
where the summations of $\alpha$ and $\beta$ are restricted to the
states within the pairing active space. 
The last term in Eq. (\ref{freeBCS2}) is a correction for
a double-counting of excitations within the pairing active space, which
stems from the substitution of 
the completeness relation $\sum_k|\phi_k\rangle\langle\phi_k| 
= 1-\sum_{\beta}|\phi_{\beta}\rangle\langle\phi_{\beta}|$ 
in Eq. (\ref{freeBCS}). 
Note that, without the pairing correlations, 
$E_{\alpha}=\lambda-\epsilon_{\alpha}$ and $v_{\alpha}=1$ for 
$\epsilon_{\alpha}< \lambda$, and the BCS free response function 
(\ref{freeBCS2}) is identical  to the corresponding response function
(\ref{free2}) in the RPA. 
With this unperturbed response
function (\ref{freeBCS2}), the QRPA response function is obtained again
by solving the Bethe-Salpeter equation (\ref{BS}) as in the RPA theory. 

\section{Continuum QRPA Excitations in $^{120}$S\mbox{n}}

We now apply the continuum QRPA formalism to 
$^{120}$Sn and make a comparison between the QRPA and the RPA. We choose 
this system because it is a typical open-shell nucleus and
also because it is a sub-shell closure nucleus where the RPA can be
applied unambiguously without using the filling approximation for
valence nucleons. 
The single-particle wave functions $\phi$ and the
single-particle energies $\epsilon$ are obtained by solving the Shr\"odinger
equation with a Woods-Saxon mean-field potential. 
As a residual interaction $v_{res}$, we use the
$t_0$ and $t_3$ parts of the Skyrme residual interaction, which is
obtained from the second derivative of the Skyrme energy functional with
respect to the proton and the neutron densities. 
The ground state density to be used in the density-dependent $t_3$ 
part of the residual interaction is generated from the single-particle 
wave functions $\phi$. 
We use the same parameters as those in Ref. \cite{SB75} 
for the mean-field potentials and the residual interaction, i.e., 
$t_0=-$1100 MeV fm$^3$, $t_3=$16000 MeV fm$^6$, 
$x_0$=0.5, $x_3$=1, and $\gamma$=1 in the standard notation of 
the Skyrme functional \cite{VB72}. 
Since this model is not self-consistent,  we renormalize 
$v_{res}$ for  the QRPA
and the RPA calculations, respectively,
so that the spurious IS dipole mode appears at zero
excitation energy. 
The resultant renormalization factors are $\kappa$ = 0.638 and 
0.660 for RPA and QRPA, respectively. 
For the pairing of neutrons, we use a schematic state-independent
constant pairing gap $\Delta_{\alpha}=\Delta=1.392$ MeV, which is 
estimated from the experimental binding energies
of neighboring nuclei. 
The pairing active space which we adopt includes the levels up to 
the $N=50$ major shell as well as the 
1g$_{7/2}$, 2d$_{5/2}$, 2d$_{3/2}$, 3s$_{1/2}$, and 1h$_{11/2}$ levels. 
The single-particle energy for the valence levels   
are shown in Table 1, together with 
the BCS occupation probabilities. 
The proton pairing gap is set to be zero 
due to the magic number $Z=50$.

Figs. 1--4 show the strength function $S(\omega)$ for the isoscalar
(IS) monopole, IS  quadrupole, IS octupole, and the isovector (IV)
dipole modes of excitations, respectively. 
The external field $V_{ext}(r)$ for each modes is given by 
$r^2$, $r^2$, $r^3$, and $r\tau_z$, 
respectively.
The lower panels show results of
the QRPA calculations, while the upper panels show results of the RPA
calculations without the pairing correlations as a comparison. For the
latter, the levels up to  3s$_{1/2}$ state are occupied with the
occupation probability $v^2$ of 1, while those  above 1h$_{11/2}$ state
 are
unoccupied with $v^2=0$ (see Table 1). 
Arrows indicate the positions of (Q)RPA states below the threshold,
where the strength function has no width.  

For the monopole excitation shown in Fig.1, the RPA does not show any 
low-lying mode, although the experimental second lowest $0^+$
state in the $^{120}$Sn nucleus 
is observed at the excitation energy 1.874 MeV, which 
is attributed to a paring vibration mode. 
The second lowest $0^+$ state in the QRPA is found at rather low energy 
2.9 MeV, and may couple to the pairing vibrational mode. 
The IS giant monopole states are seen in the energy region 17$-$28
MeV in both the QRPA and RPA calculations.  
The transition densities for the IS monopole mode at
two different energies, i.e., $\omega$ = 2.9 MeV and 21.6 MeV, 
are shown in Fig. 5. 
We find that the former shows 
a characteristic behaviour of the pairing vibration mode while the
latter shows the compressional character. 
For the quadrupole mode shown in Fig. 2, the lowest RPA state is at
5.2 MeV, in comparison with the experimental value of 1.17 MeV. 
The transition strength is B(E2:0$^+\to2^+$)= 0.031 $e^2b^2$ in the 
RPA calculation, while the experimental value is 
0.2 $e^2b^2$. 
If one takes the pairing correlations into account,  the 
low-lying 2$^+$ RPA states  goes  substantially  down 
and the lowest QRPA state appears at 2.3 MeV having the transition
strength B(E2:0$^+\to2^+$)= 0.107 $e^2b^2$, which is much closer to
the experimental value. In both cases, the main peak of 
the IS giant quadrupole resonance (GQR) 
   appears at around 13 MeV exhausting most
 of the sum rule strength.  The experimental GQR is also observed at
the excitation energy around 13MeV\cite{Woude80}. 
As for the octupole mode of excitations in Fig. 3, the experimental
value of the lowest $3^-$ state is at 2.4 MeV, while the RPA and the
QRPA lead to 1.4 MeV and 3.0 MeV, respectively. 
The experimental value for the corresponding transition rate is 
B(E3:0$^+\to3^-$)= 0.09 $e^2b^3$, while it is 0.159 $e^2b^3$ and 
0.0771 $e^2b^3$ in the RPA and the QRPA, respectively. 
The RPA without pairing correlations underestimates the
lowest $3^-$ state energy and the QRPA is more satisfactory
compared with the experimental value. 
We summarize in Table 2 the results of the RPA and the 
QRPA calculations 
together with the experimental data for the lowest 2$^+$ and 3$^-$
states. 
Giant octupole resonances (GOR) are
observed in Fig. 3 at the excitation energy 23 MeV in both cases.
Results for the IV giant dipole resonance (GDR) are 
shown for RPA and QRPA in Fig. 4. 
As is the same as in Figs. 1--3, the  structure of GDR is not much disturbed 
by the effect of the pairing.

Notice that the 
effect of the pairing on the low-lying quadrupole
mode is opposite to that on the octupole mode. For the former, the
energy of the lowest 2$^+$ state becomes smaller due to the pairing,
with enhancement of the B(E2) value. On the other hand, the QRPA raises the
energy of the lowest 3$^-$ state and the B(E3) value is hindered. 
In both cases, the QRPA gives better agreement with the experimental
values compared with the RPA. 
These features can be understood as follows. 
For excitations with even parity, such as the quadrupole excitations,
transitions between the same levels are allowed in the presence of the
pairing correlations. This enhances the
collectivity of low-lying states, resulting in a smaller excitation
energy and a larger transition strength. On the contrary, such
excitations are not allowed for odd parity excitations, e.g., the
octupole mode. In that case, the dominant excitation is from one major 
shell to another regardless of the pairing correlation. Since the
particle-hole excitations are weakened by the factor 
$v^2$ in the presence of the
pairing, the QRPA lowers the collectivity of low-lying collective
excitations with odd parity. 

In contrast to the low-lying modes of excitations, in general, 
high-lying modes are much less sensitive to the paring correlations. This is
the case for all the modes of excitations shown in Figs. 1--4. 
This fact can be seen more transparently  
if one smears the strength functions
with a finite width. We show in Figs. 6--9 the RPA and QRPA results 
with a finite value of $\eta$ =0.5MeV in Eq. (\ref{freeBCS2}). 
As one clearly sees, the two results resemble each other in the giant 
resonance region at energies
above the threshold near 10 MeV. 
These results are  natural consequences of the pairing correlations
  since the configurations  outside  the
pairing active space are  the main p-h configurations for  the
giant resonances and thus the pairing effects should play a minor
role. 

 A conventional  approximation to treat the
continuum effect is to put a nucleus in a box and discretize the
continuum states by imposing a boundary condition  at the edge of the box.
A model space of these states is usually truncated at the 
maximum single-particle energy $\epsilon_{max}$. 
Since our continuum QRPA method can treat the coupling to the continuum 
exactly, it is interesting  to compare our  method with the box discretization
 approximation. Fig.  10 shows a convergence property
of the box discretization method for the IS quadrupole response. 
We take the box size of $R_{max}$=10 fm, and smear the strength
function with $\eta$=0.5 MeV for the representation purpose. 
We have checked that the results do not change significantly even if
we use a larger value of $R_{max}$. 
The solid line is the result of the
continuum QRPA method with the exact treatment of the continuum
effect, which is the same as in Fig. 6. 
The dot-dashed, dashed, and thin solid lines are results of the box
discretization method with truncation energy at $\epsilon_{max}$= 10, 50, and
100 MeV, respectively. In the upper panel, we use the same residual
interaction for all the calculations. As one can see, the convergence
with respect to the maximum energy of continuum states is extremely
slow: even with $\epsilon_{max}$= 100 MeV the peak positions are not
reproduced. The peak height for the lowest energy peak is not reproduced,
either. 
We note that the truncation of the model space shfts the position of 
the spurious IS dipole state. It appears at 6.0, 4.4, and 3.7 MeV for 
$\epsilon_{max}$= 10, 50, and 100 MeV, respectively. 
In the lower panel,  the renormalized
residual interaction is used for each value of $\epsilon_{max}$ 
so that the IS
dipole mode appears at zero energy. 
The renormalization factors are $\kappa$= 0.968, 0.776, and
0.732 for $\epsilon_{max}$= 10, 50, and 100 MeV,
respectively. 
This procedure drastically
improves the convergence. 
In the case of the truncation energy $\epsilon_{max}$= 50 MeV (dashed line),
the exact solution is almost reproduced. The result converges
at this energy and the shape of the strength function is not altered
even when the  states up to $\epsilon_{max}$= 100 MeV are
included. This study clearly indicates the importance of
 the self-consistency between the space truncation and the renormalization of
the residual interaction. If  this self-consistency
is imposed,  the box discretization
method works well. It should be emphasized, however, that the escape
width of resonances can be obtained only when the
single-particle continuum states are treated exactly as we do in this article. 

\section{Summary and Discussions}

We proposed a QRPA model in the coordinate space to take into account
the continuum effects. 
It is a generalization of 
the formalism of Shlomo and Bertsch for the continuum 
RPA to the QRPA. We treated separately the p-h excitations within the pairing 
active space and those between the active space and the non-active
space. 
For the former, we explicitly used the two quasiparticle configurations in
the response function in the coordinate space, 
while for the latter we 
use the single-particle Green function in the coordinate space 
representation taking into account the coupling to 
the particle continuum  properly. 
We applied the formalism to  $^{120}$Sn and showed that 
the pairing correlations enhance the collectivity of the positive
parity low-lying states, while that of the negative parity low-lying 
states are hindered. We found that 
the QRPA is more satisfactory in reproducing the experimental data of 
the energy of low-lying modes of excitation, while the giant
resonances are not much affected by the pairing correlations. 
We would like to point out that these results  however do not 
 justify the
RPA model without the pairing correlations for the  study of
 giant resonances in 
open-shell nuclei, especially near the drip-lines. 
Since the  single-particle energies and wave functions
could be   different 
in principle due to the density dependence of the mean-field,  
the  response properties even in the giant resonance region might
be affected by the pairing correlations. 
We therefore advocate to use the QRPA all through the excitation
region in open-shell nuclei. 

There are many possible applications of the continuum QRPA method
presented in this paper. One of them is a neutrino-nucleus scattering,
e.g. $^{12}$C($\nu,\nu'$)$^{12}$C$^*$ \cite{JRHR99} and 
$^{12}$C($\nu_e,e^-$)$^{12}$N \cite{VAC00}, where the RPA and the
QRPA have been one of the standard methods for theoretical investigations. 
Another application will be the excitations of drip-line nuclei. Because of 
a low energy threshold, the responses would be very 
sensitive to the pairing correlations. In these cases, the
pairing active space includes particle continuum
states and the BCS approximation 
should be carefully examined. 
Also, the residual interaction in the particle-particle 
channel, which we neglected in this paper, might have to be taken 
into account. 
A more sophisticated  treatment of the pairing interaction is
provided by the Hartree-Fock-Bogoliubov (HFB) theory, 
which consistently describes couplings between the
particle-hole and the particle-particle channels.
It would be an
interesting future work to develop a continuum QRPA theory based on
the HFB approximation. 

\section*{Acknowledgment}

We are grateful to the Institute for Nuclear Theory at the University of
Washington, where this work was initiated during the INT program INT-00-3 
on ``Nuclear structure for the 21st century''. We thank its
hospitality and a partial financial support for this work. 
We thank also K. Matsuyanagi and Nguyen Van Giai 
for useful and stimulating discussions and Baha
Balantekin for discussions on the RPA approach to neutrino-nucleus reactions. 
This work is supported in part by the Ministry of Education, 
Science, Sports and Culture in Japan by Grant-In-Aid for Scientific
Research under the program number (C(2)) 12640284.

\newpage

\begin{table}[hbt]
\caption{Neutron single-particle levels near the Fermi surface 
for $^{120}$Sn obtained with a Woods-Saxon potential. 
The occupation probabilities $v^2$ are
calculated in the BCS approximation with a constant pairing gap
$\Delta=$ 
1.392 MeV. The pairing active space includes those levels shown in
this table as well as the levels up to the $N=50$ major shell. 
The neutron Fermi energy $\lambda_n$ is $-$8.149 MeV. }

\vspace*{-2pt}
\begin{center}

\begin{tabular}{ccc}

level & energy (MeV) & occupation probability \\
\hline

1g$_{7/2}$ & $-$11.184  &   0.954 \\
2d$_{5/2}$ & $-$11.145  &   0.953 \\
2d$_{3/2}$ & $-$9.280   &   0.815 \\
3s$_{1/2}$ & $-$9.048   &   0.771 \\
1h$_{11/2}$ & $-$6.949 &    0.173 

\end{tabular}
\end{center}

\end{table}

\begin{table}[hbt]
\caption{Comparison of the RPA and the QRPA calculations with the 
experimental data for the lowest-lying 2$^+$ and 3$^-$ modes of
excitation of $^{120}$Sn.}

\vspace*{-2pt}
\begin{center}

\begin{tabular}{c|cc|cc}

 & E$_{2^+}$ (MeV) & B(E2)(e$^2$b$^2$)
&E$_{3^-}$ (MeV) & B(E3)(e$^2$b$^3$)  \\
\hline
RPA & 5.2  &   0.031 & 1.4 & 0.159\\
QRPA & 2.3 &   0.105 & 3.0 & 0.0771 \\
\hline
Expt & 1.17 &   0.2 & 2.4 & 0.09

\end{tabular}
\end{center}

\end{table}

\newpage

\begin{figure}
  \begin{center}
    \leavevmode
    \parbox{0.9\textwidth}
           {\psfig{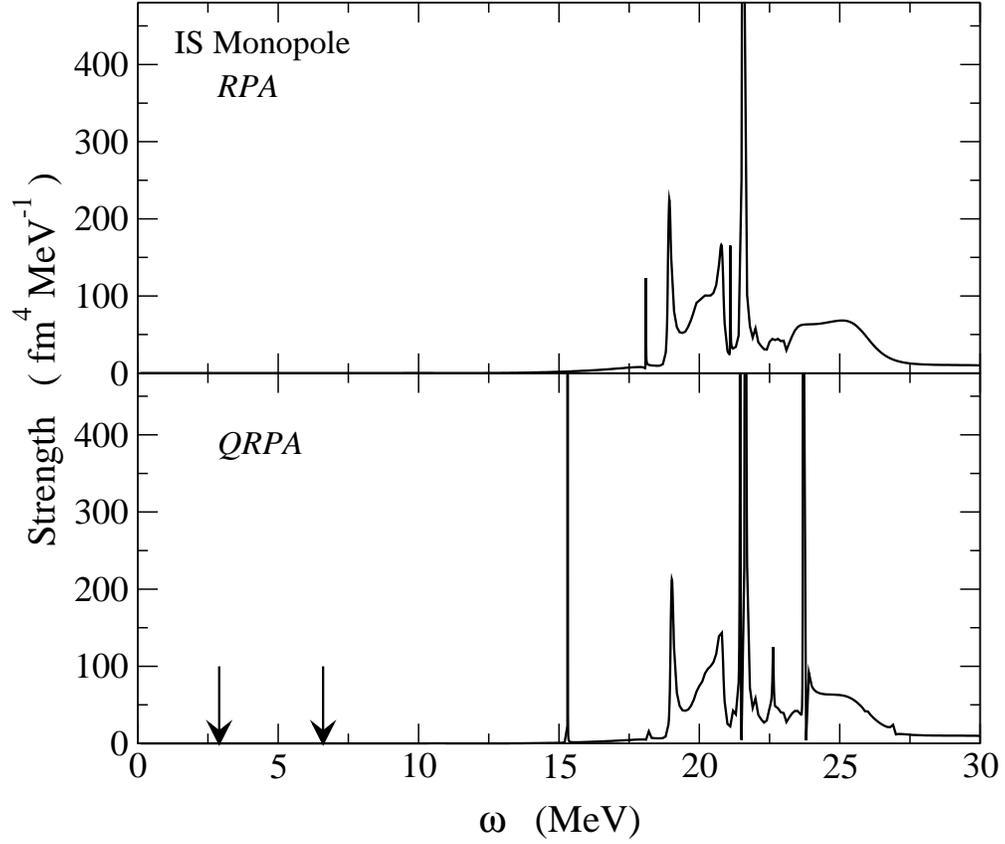}}
  \end{center}
\protect\caption{
Comparison between the RPA (the upper panel) and the QRPA (the lower
    panel) for the isoscalar monopole mode in the $^{120}$Sn nucleus. 
Arrows indicate the position of QRPA states below the threshold. }
\end{figure}

\newpage

\begin{figure}
  \begin{center}
    \leavevmode
    \parbox{0.9\textwidth}
           {\psfig{file=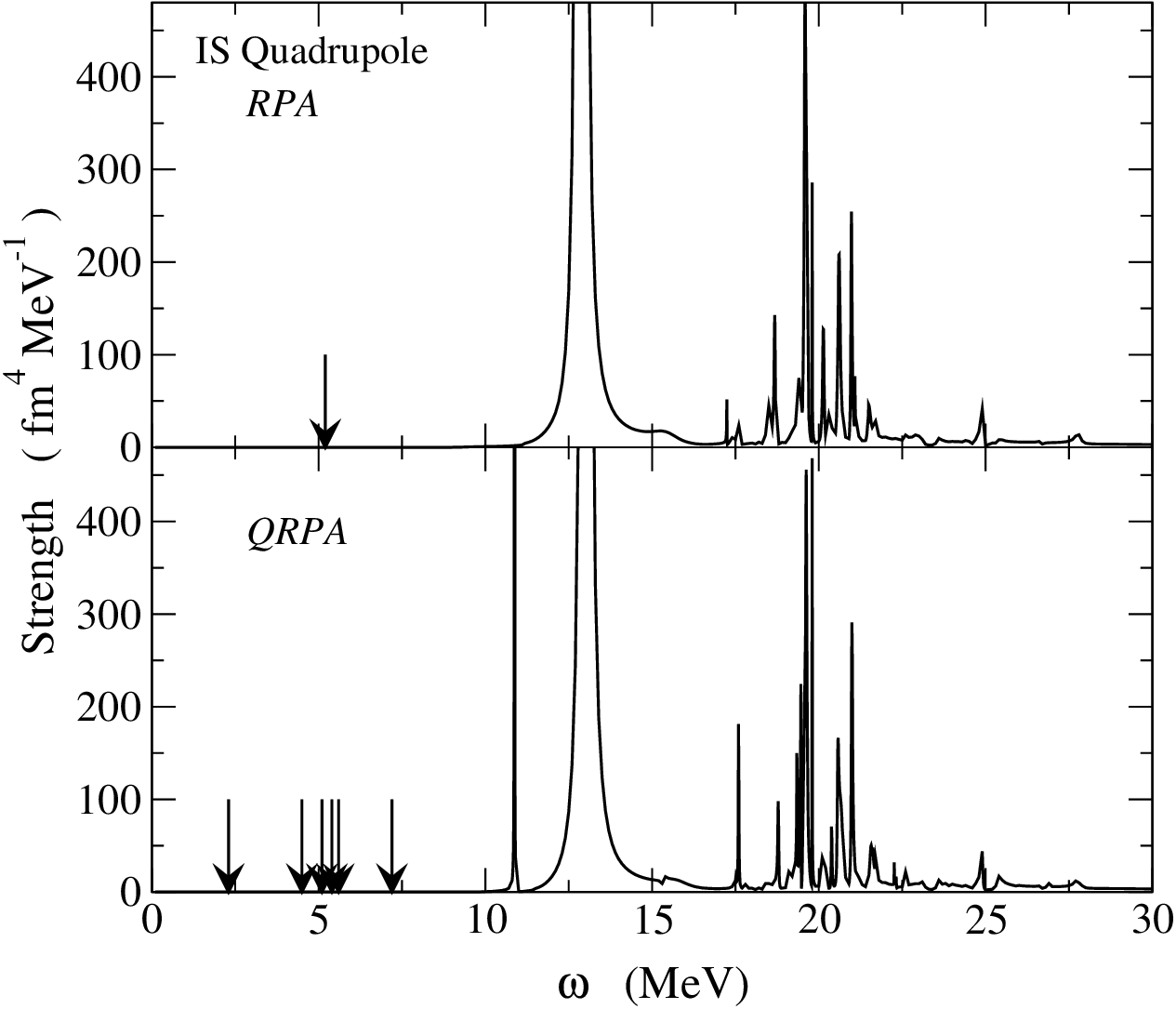,width=0.8\textwidth}}
  \end{center}
\protect\caption{Same as fig. 1, but for the isoscalar quadrupole mode. }
\end{figure}

\newpage

\begin{figure}
  \begin{center}
    \leavevmode
    \parbox{0.9\textwidth}
           {\psfig{file=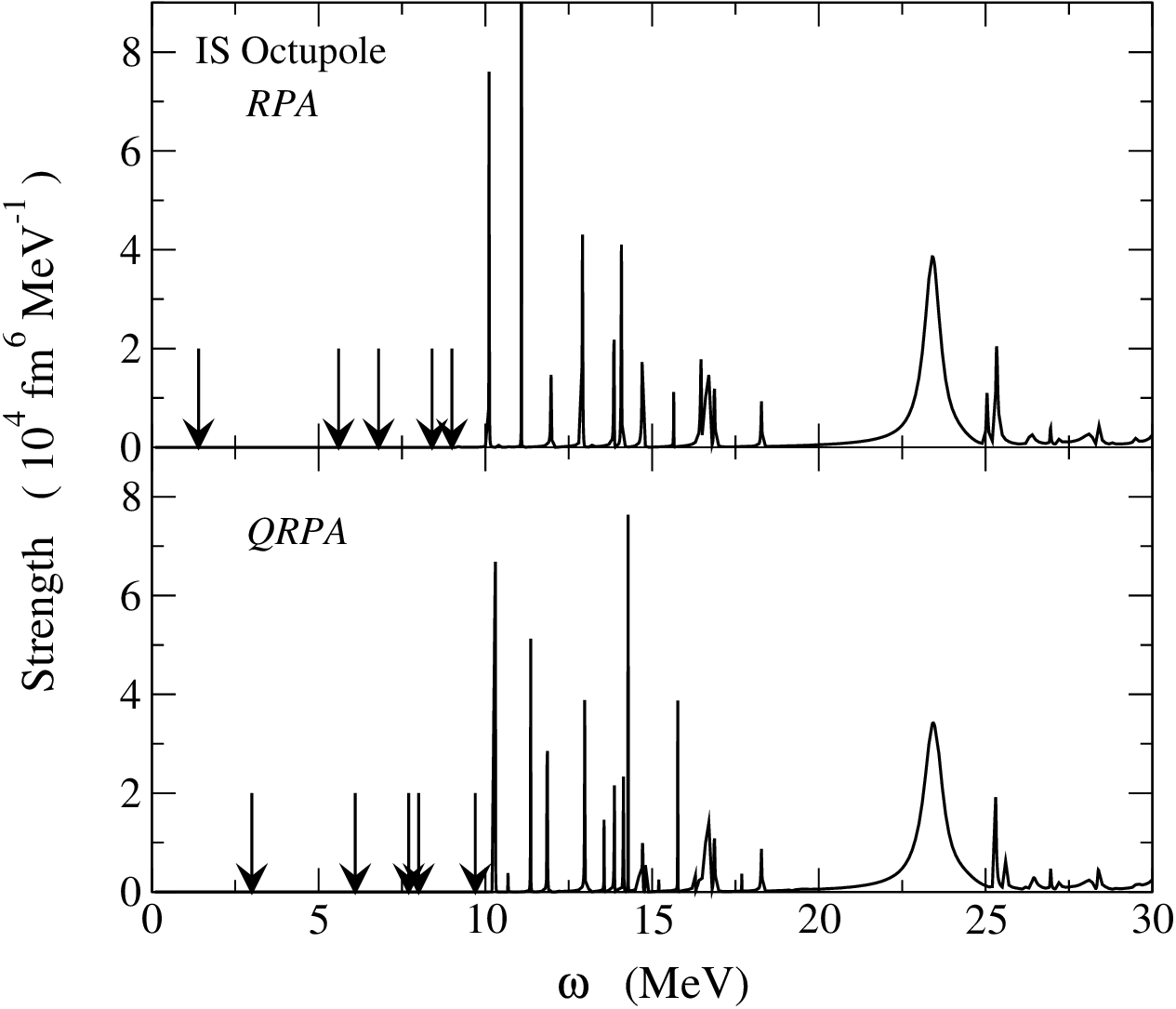,width=0.8\textwidth}}
  \end{center}
\protect\caption{Same as fig. 1, but for the isoscalar octupole mode. }
\end{figure}

\newpage
\begin{figure}
  \begin{center}
    \leavevmode
    \parbox{0.9\textwidth}
           {\psfig{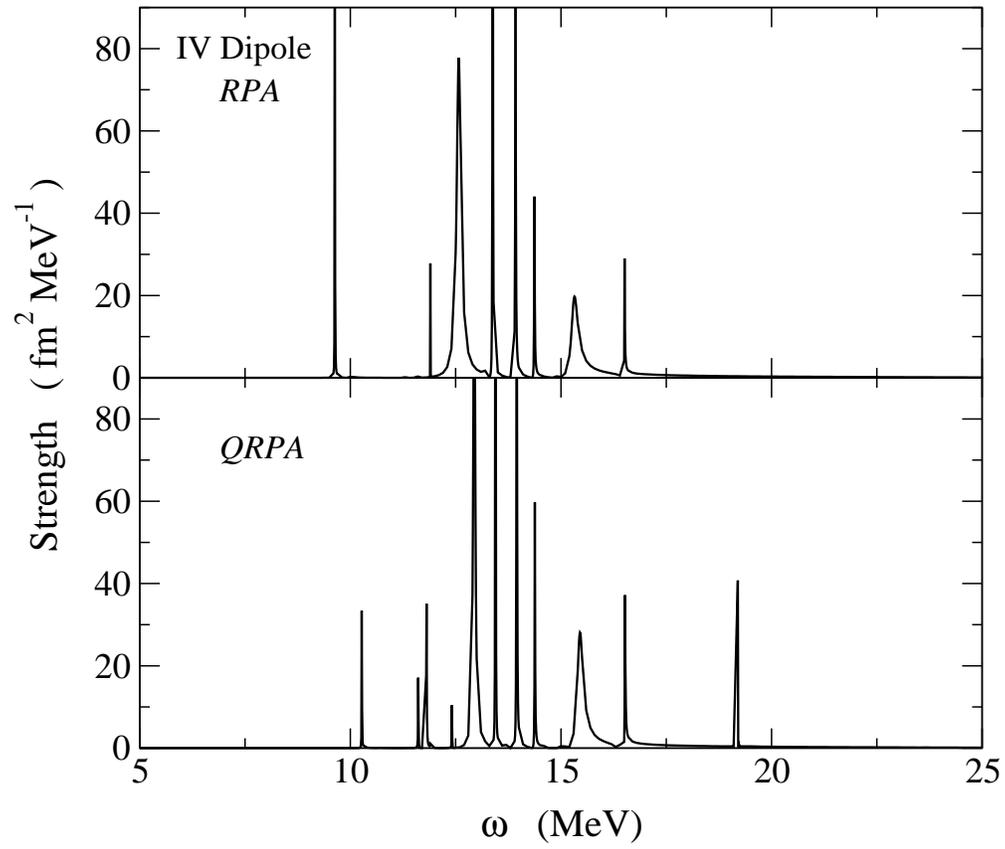}}
  \end{center}
\protect\caption{Same as fig. 1, but for the isovector dipole mode. }
\end{figure}

\newpage

\begin{figure}
  \begin{center}
    \leavevmode
    \parbox{0.9\textwidth}
           {\psfig{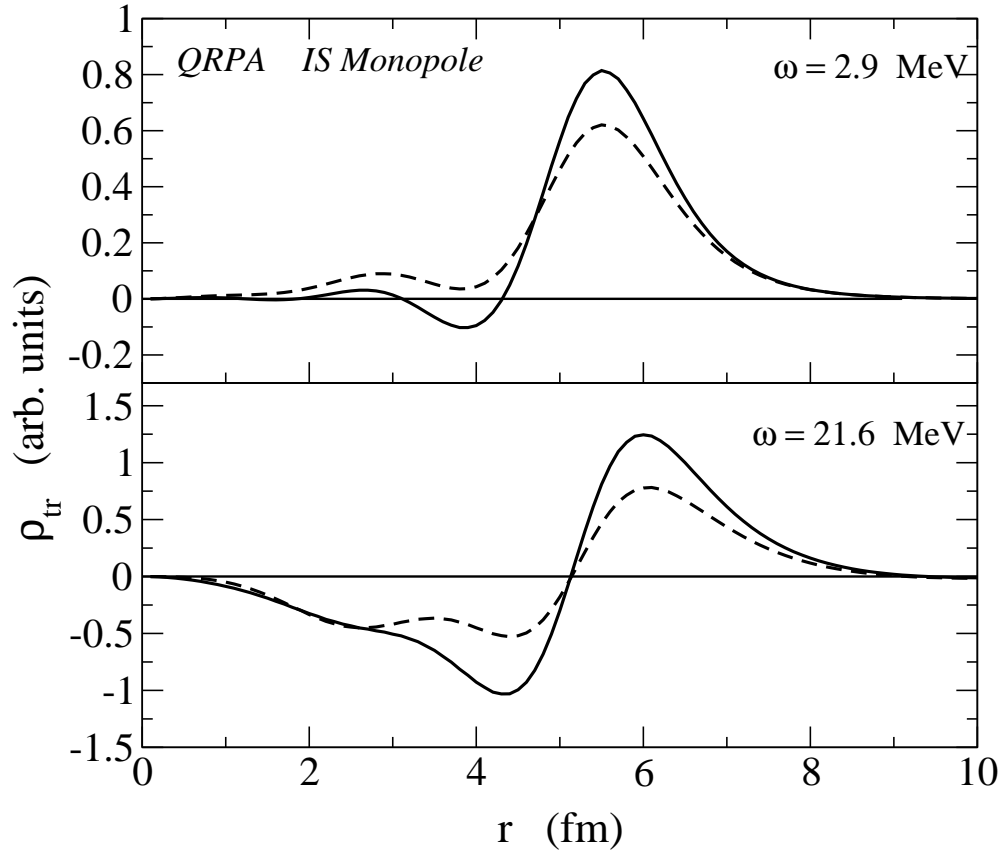}}
  \end{center}
\protect\caption{Transition densities for the isoscalar monopole mode
    obtained in the QRPA (the solid line). The upper and the lower
panels are for
    $\omega$ = 2.9 and 21.6 MeV, respectively. The neutron
contribution is plotted separately by the dashed line. }
\end{figure}

\newpage

\begin{figure}
  \begin{center}
    \leavevmode
    \parbox{0.9\textwidth}
           {\psfig{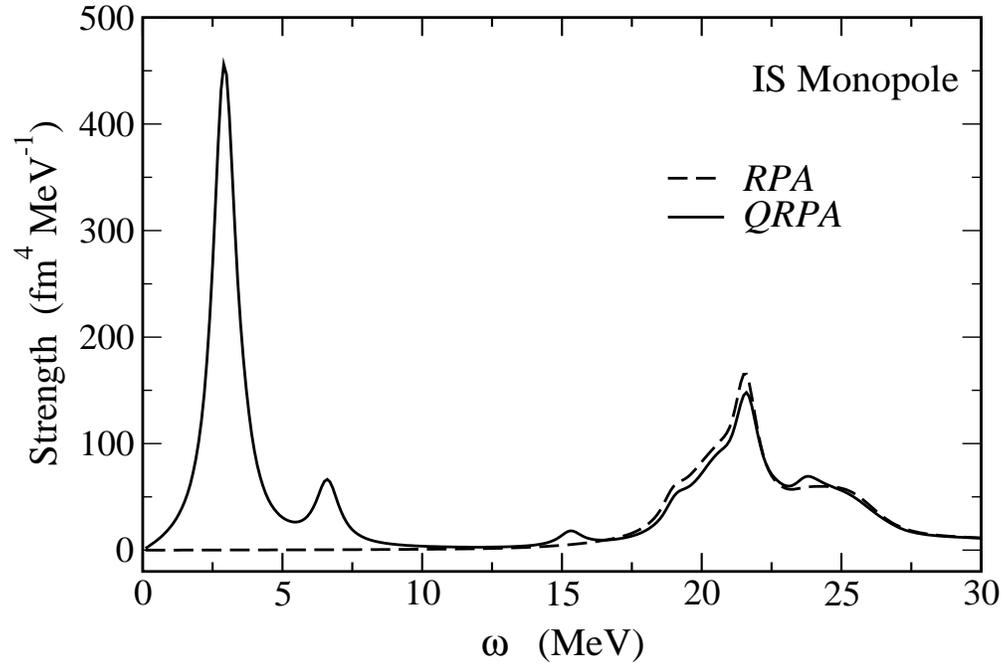}}
  \end{center}
\protect\caption{Same as fig.1, but with $\eta$=0.5 MeV in 
Eq. (\ref{freeBCS2}), which is equivalent to smearing the strength
    function with the width of 1 MeV. 
}
\end{figure}

\newpage

\begin{figure}
  \begin{center}
    \leavevmode
    \parbox{0.9\textwidth}
           {\psfig{file=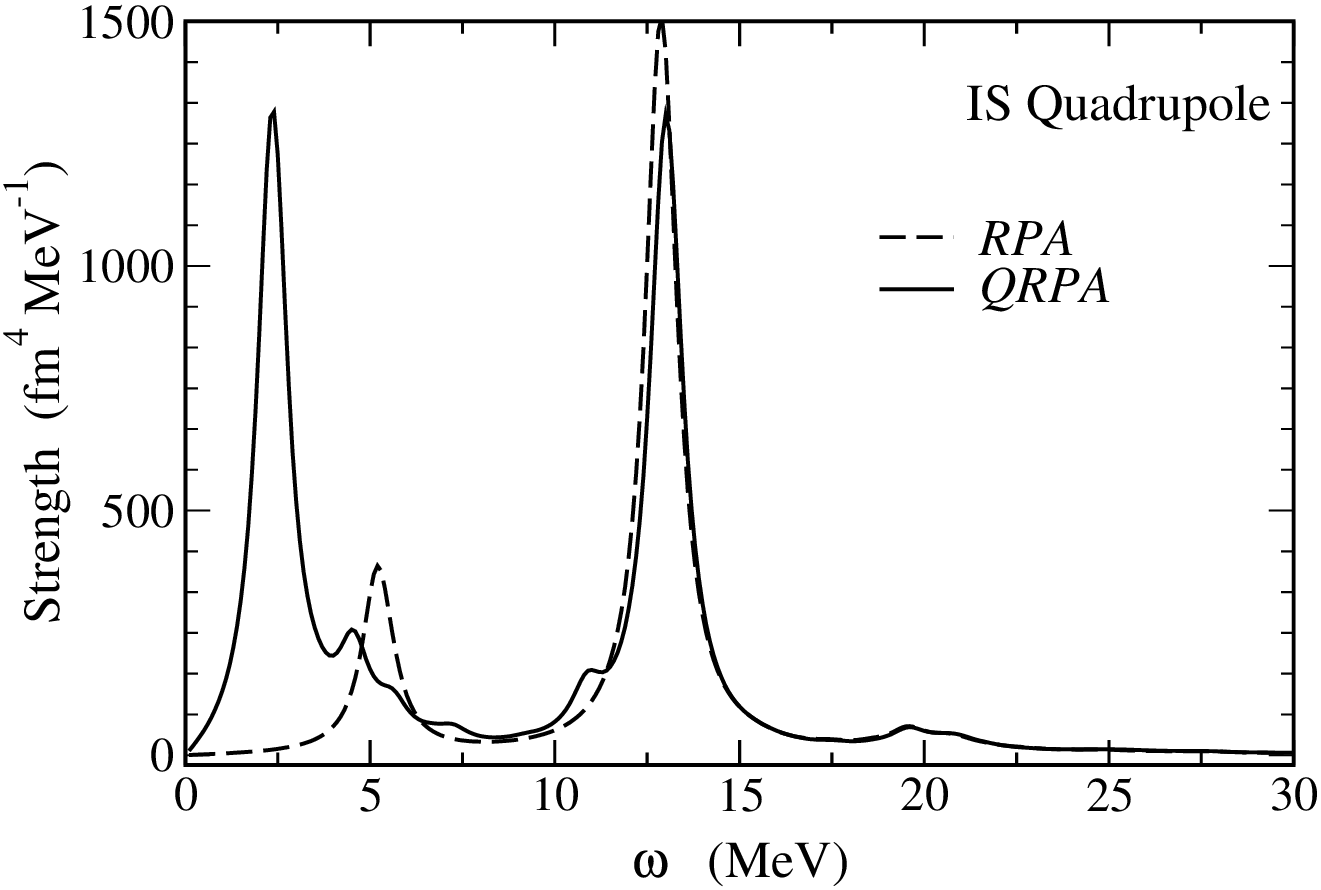,width=0.8\textwidth}}
  \end{center}
\protect\caption{Same as fig. 6, but for the isoscalar quadrupole mode. }
\end{figure}

\newpage

\begin{figure}
  \begin{center}
    \leavevmode
    \parbox{0.9\textwidth}
           {\psfig{file=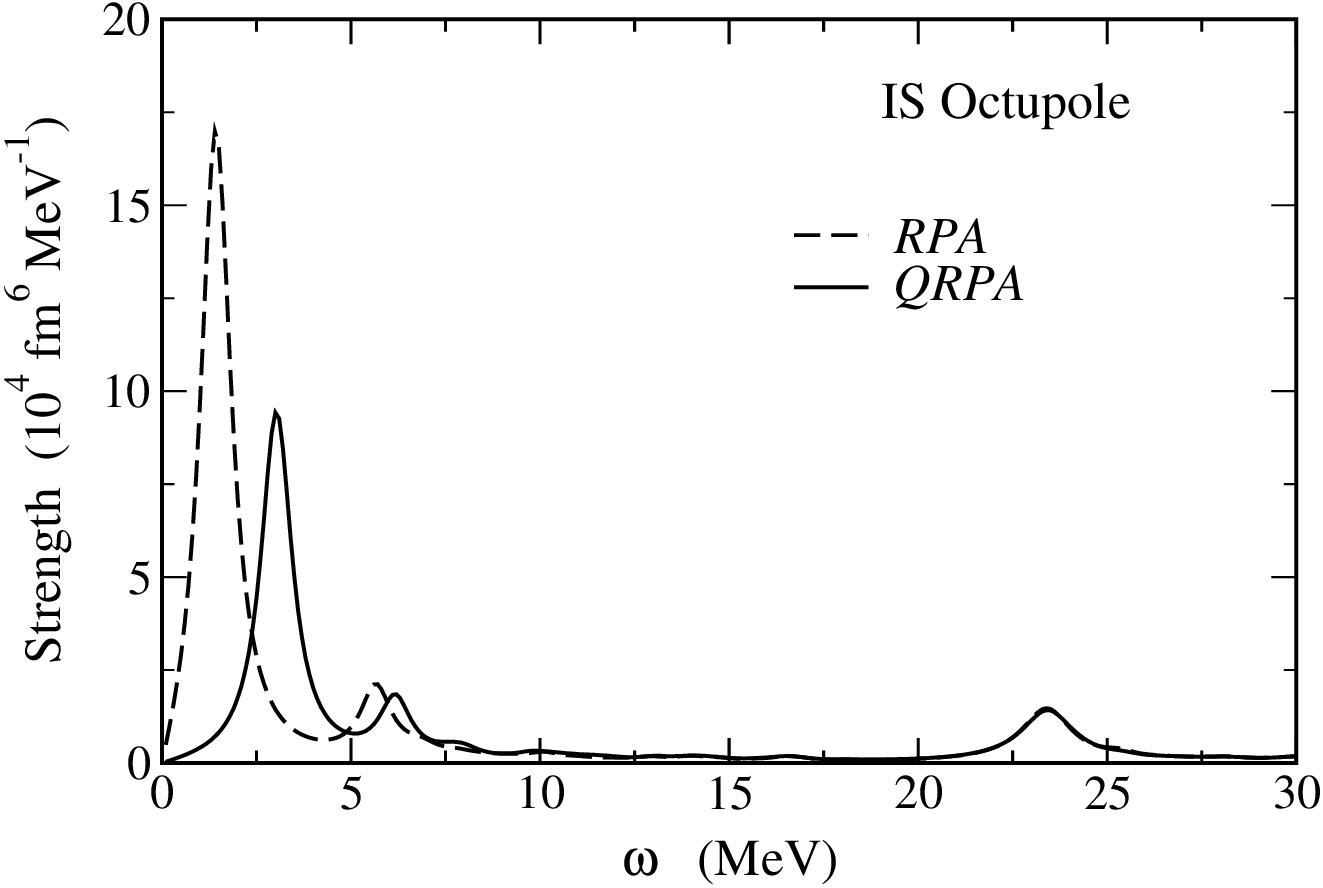,width=0.8\textwidth}}
  \end{center}
\protect\caption{Same as fig. 6, but for the isoscalar octupole mode. }
\end{figure}

\newpage

\begin{figure}
  \begin{center}
    \leavevmode
    \parbox{0.9\textwidth}
           {\psfig{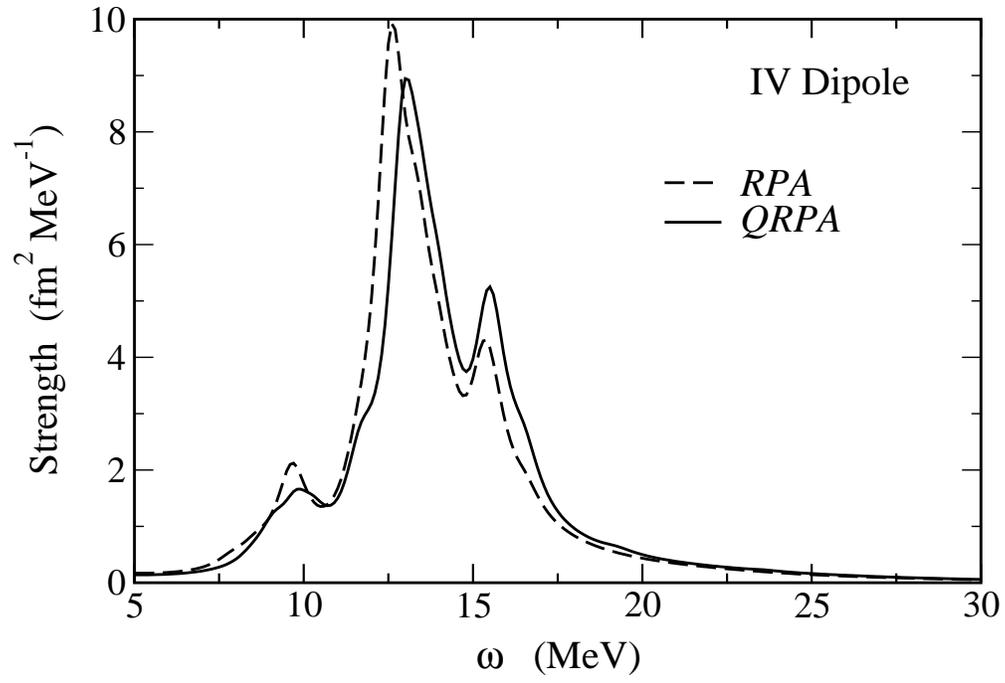}}
  \end{center}
\protect\caption{Same as fig. 6, but for the isovector dipole mode. }
\end{figure}

\newpage

\begin{figure}
  \begin{center}
    \leavevmode
    \parbox{0.9\textwidth}
           {\psfig{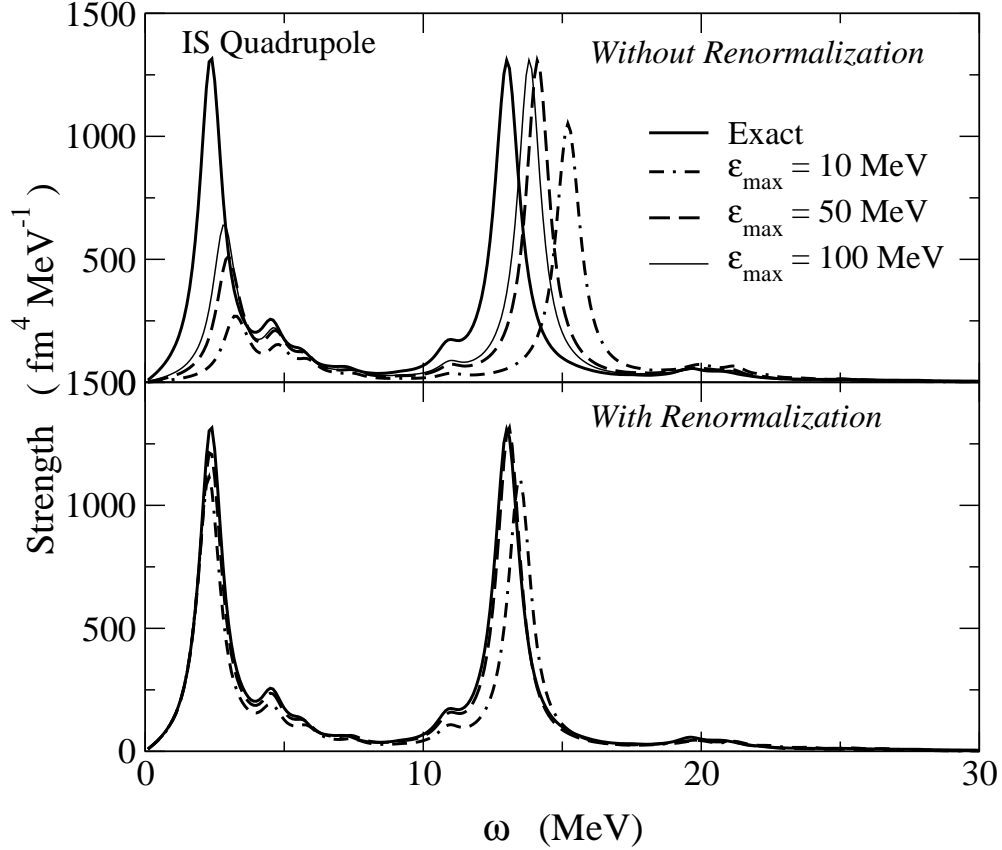}}
  \end{center}
\protect\caption{Convergence property of the box discretization method
    for the continuum coupling 
for the isoscalar quadrupole mode of $^{120}$Sn. 
The box size for the discretization is taken to be 10 fm, and results are
    smeard with $\eta$=0.5 MeV. The thick solid line is the result of
    the continuum QRPA calculations which treat the continuum effects
    exactly. The upper and lower panels are obtained without and with
    imposing the self-consistency condition 
of the residual interaction, respectively.}
\end{figure}


\begin{references}

\bibitem{BB94}G.F. Bertsch and R.A. Broglia, 
{\it Oscillations in Finite Quantum Systems} (Cambridge University 
Press, Cambridge, 1994). 

\bibitem{FW71} A.L. Fetter and J.D. Walecka, {\it Quantum Theory 
of Many-Particle Systems} (McGraw-Hill, New York, 1971). 

\bibitem{R68}D.J. Rowe, {\it Nuclear Collective Motion} (Methuen,
London, 1968). 

\bibitem{RS80}
P. Ring and P. Schuck, {\it The Nuclear Many Body Problem} 
(Springer-Verlag, New York, 1980).

\bibitem{SB75} S. Shlomo and G. Bertsch, Nucl. Phys. {\bf A243}, 
507 (1975). 

\bibitem{RG92}P.-G. Reinhard and Y.K. Gambhir, 
Annalen der Physik {\bf 1}, 598 (1992). 

\bibitem{BT75}G.F. Bertsch and S.F. Tsai, Phys. Rep. {\bf 18}, 
126 (1975). 

\bibitem{Tsai78}
S.F. Tsai, Phys.Rev. C{\bf 17}, 1862 (1978).

\bibitem{GSV98}Nguyen Van Giai, Ch. Stoyanov, and V.V. Voronov, 
Phys. Rev. C{\bf 57}, 1204 (1998). 

\bibitem{LG76}K.F. Liu and Nguyen Van Giai, Phys. Lett. {\bf 65B}, 
23 (1976). 

\bibitem{HSZ97}I. Hamamoto, H. Sagawa, and X.Z. Zhang, 
Phys. Rev. C{\bf 55}, 2361 (1997); 
Phys. Rev. C{\bf 57}, R1064 (1998). 

\bibitem{HS99}I. Hamamoto and H. Sagawa, Phys. Rev. C{\bf 60}, 
064314 (1999); Phys. Rev. C{\bf 62}, 024319 (2000). 

\bibitem{NY00}T. Nakatsukasa and K. Yabana, 
e-print: physics/0010025. 

\bibitem{BE91}G.F. Bertsch and H. Esbensen, Ann. of Phys. (N.Y.) 
{\bf 209}, 327 (1991). 

\bibitem{EB92}H. Esbensen and G.F. Bertsch, Nucl. Phys. {\bf A542}, 
310 (1992). 

\bibitem{KLLT98}S. Kamerdzhiev, R.J. Liotta, E. Litvinova, and 
V. Tselyaev, Phys. Rev. C{\bf 58}, 172 (1998).

\bibitem{KG00}E. Khan and Nguyen Van Giai, Phys. Lett. {\bf B472}, 253
(2000). 

\bibitem{DNW96}J. Dobaczewski, W. Nazarewicz, T.R. Werner,
J.F. Berger, C.R. Chinn, and J. Decharg\'e, Phys. Rev. C{\bf 53}, 2809
(1996). 

\bibitem{FTTZ00}S.A. Fayans, S.V. Tolokonnikov, E.L. Trykov, and 
D. Zawischa, Nucl. Phys. {\bf A676}, 49 (2000).

\bibitem{GGLS00}Nguyen Van Giai, M. Grasso, R.J. Liotta, and
N. Sandulescu, e-print: nucl-th/0010022. 

\bibitem{VB72}D. Vautherin and D.M. Brink, Phys. Rev. C{\bf 5}, 626 (1972).

\bibitem{Woude80}
 A. van der Woude, 
 Proc. of Giant Multipole Resonance Topical Conference (ed. F. E. Bertrand,
 Oak Rigde,  1979)  p. 65

\bibitem{JRHR99}N. Jachowicz, S. Rombouts, K. Heyde, and
J. Ryckebusch, Phys. Rev. C{\bf 59}, 3246 (1999). 

\bibitem{VAC00}
C. Volpe, N. Auerbach, G. Col\'o, T. Suzuki, and N. Van Giai, 
Phys. Rev. C{\bf 62}, 015501 (2000). 

\end{references}
\end{document}